\documentclass[usenatbib]{mnras}

\usepackage[T1]{fontenc}
\usepackage{ae,aecompl}

\usepackage{graphicx}	
\usepackage{amsmath}	
\usepackage{amssymb}	

\title[Modelling GRB host galaxy populations with IllustrisTNG]{One star, two stars, or both? Investigating metallicity-dependant models for Gamma-Ray Burst progenitors with the IllustrisTNG simulation}

\author[Metha, Trenti]{
Benjamin Metha$^{1}$\thanks{\hbox{methab@student.unimelb.edu.au,} mtrenti@unimelb.edu.au}
and Michele Trenti$^{1,2}$
\\
$^1$School of Physics, The University of Melbourne, VIC 3010, Australia\\
$^2$Australian Research Council Centre of Excellence for All-Sky Astrophysics in 3-Dimensions, Australia
}

\date{Accepted XXX. Received YYY; in original form ZZZ}

\pubyear{2019}

\begin{document}
\label{firstpage}
\pagerange{\pageref{firstpage}--\pageref{lastpage}}
\maketitle

\begin{abstract}
The rate of long-duration gamma ray bursts (GRBs) has been identified as a potential proxy for the star formation rate (SFR) across redshift, but the exact relationship depends on GRB progenitor models (single versus binary). The single-progenitor  collapsar model accounts for the preference towards low-metallicity GRB progenitors, but is in apparent tension with some high-metallicity GRB host galaxy measurements. As a possible solution, we consider the scenario where high-metallicity GRB hosts harbour low metallicity regions in which GRB progenitors form. For this, we use the IllustrisTNG cosmological hydrodynamical simulation to investigate the internal metallicity distribution of GRB hosts, implementing in post-processing different GRB formation models. Predictions (GRB rate, host metallicities and stellar masses) are compared to the high-completeness GRB legacy surveys BAT6 and SHOALS and a sample of high-redshift GRB-DLA metallicities, allowing us to compute their relative likelihoods. When the internal metallicity distribution of galaxies is ignored, the best-fitting model requires a metallicity-independent channel, as previously proposed by Trenti, Perna, \& Jimenez. However, when the internal metallicity distribution is considered, a basic metallicity bias model with a cutoff at $Z_{max}=0.35Z_\odot$ is the best fitting one. Current data are insufficient to discriminate among more realistic metallicity bias models, such as weak metallicity dependence of massive binaries vs stronger metallicity bias of collapsars. 
An increased sample of objects, and direct measurements of host stellar masses at redshift $z>2$ would allow to further constrain the origin of long GRBs.
\end{abstract}



\section{Introduction} \label{sec:intro}

Gamma ray bursts (GRBs) are the brightest explosions observed in the universe. With energies of the order $10^{53}$ ergs, these bursts typically last for a few seconds, before they decay into longer lived afterglows across the electromagnetic spectrum, including rest-frame ultraviolet to infrared. GRBs are divided into two categories: short GRBs, which release $90\%$ of their energy in the first $2$ seconds, and long GRBs, which take longer than $2$ seconds to release $90\%$ of their energy \citep{Two_classes}. Because of their extreme luminosities, GRBs offer a unique way to see star-forming galaxies at high redshift that are too small or faint to be observed directly \citep{Chary+07, Fynbo+08, Griener+15, Chary+16}.

The coincident discovery of GRB170817A with GW170817 showed that at least some short GRBs are produced during merger events of neutron star pairs \citep{LIGO_GRB}; however, the physical origin of long GRBs\footnote{Hereafter we will refer to long GRBs simply as GRBs.} is still under debate. Tens of GRBs have been associated with Type Ic supernovae counterparts \citep{GRB=SNe, GRB=SNe2, Cano13}, indicating that the formation of GRBs is connected to the death of massive stars, and so the rate of production of GRBs should trace the star formation rate (SFR) of their host galaxies to first approximation (e.g. see \citealt{Jakobsson+05,Robertson+Ellis12, trenti13}, \mbox{\citealt{Chary+16, Lyman+17}}). \citet{Woosley93} first developed a theoretical framework qualitatively consistent with these data, based on the idea that a GRB could be formed from the core collapse supernova of a single, fast-spinning star (the ``collapsar" model). A few years later, inspiral and subsequent collapse of a binary system of two massive stars was proposed as an alternative model \citep{Fryer+Heger05, Cantiello+07}. 

These two models differ in the metallicity ($Z$) requirements of the progenitor system. Stars with higher metallicity have stronger stellar winds, and lose a larger fraction of their angular momentum before collapsing, suppressing their likelihood of creating a GRB on collapse \citep{Vink01}. For this reason, in the single-star progenitor model it is expected that the production of GRBs should be restricted to galaxies with lower metallicities. Conversely, in the binary progenitor based model, the angular momentum necessary to produce a GRB is transferred from the orbits of the system into stellar rotation, and so a much less severe metallicity bias is expected \citep{Chrimes+2020}.

Using hydrodynamic stellar simulations, \citet{Yoon+06} derived the functional form of the metallicity bias for the single-star collapsar model, investigating stars with four different initial metallicities, and a wide range of different initial masses and spins.
They found that the lower the metallicity of a star, the more likely it is to form a GRB on collapse, and that stars with $Z \gtrsim 0.3Z_\odot$ should never produce GRBs.

Building on this work, \citet{Chrimes+2020} computed the metallicity bias for GRB progenitors originating from binary systems using the stellar evolution code BPASS \citep{Eldridge+17}. They found that when tidal interactions were implemented, GRBs could still be generated (with vanishing probability) from systems with metallicities as high as $1.1Z_\odot$, with GRB activity peaking at $\sim 0.2Z_\odot$.

Observationally, many studies of GRB host galaxies have found that GRBs are less likely to form in high metallicity environments \citep{Fynbo+03, Modjaz+08, gehrels09, Salvaterra+12, Boissier+13, Graham&Fruchter13, Perley+13, Vergani+15,Graham&Fruchter2017, Palmerio+2019}. Recent work by \citet{Vergani+17} confirmed evidence of a mild metallicity bias, with a threshold metallicity of  $Z_{\mathrm{max}}=0.73_{-0.07}^{+0.08} Z_{\odot}$ -  significantly higher than the predicted metallicity cutoff from the single-star progenitor model.
Furthermore, there have been observations of GRB host galaxies with supersolar metallicities $Z \gtrsim Z_\odot$ (e.g. see \citealt{levesque10, Savaglio+12, Heintz+18, Michalowski+18}). These observations have prompted some studies to rule out the possibility that the collapsar model is the sole mechanism  responsible for the formation of GRBs (e.g. \citealt{Trenti+15, Chrimes+2020}). However, there are several other factors, often ignored, that could explain these higher metallicity hosts even from a model that includes a sharp metallicity cutoff.

Firstly, most surveys derive the metallicity of high-redshift GRB host galaxies using optical emission-line diagnostics, which trace the abundance of oxygen through star-forming regions. Recently, \citet{Hashimoto+18} highlighted that many young galaxies may be rich in oxygen, yet poor in iron, the abundance of which determines the strength of stellar winds \citep{Yoon+06}. For this reason, many young, star forming GRB host galaxies may have an $O/Fe$ overabundance, and relying on oxygen abundances alone could ``mislead us on the origins of GRBs".

Secondly, there is no reason to have only one of these mechanisms responsible for the production of all GRBs. \citet{Trenti+15} developed a framework that combines Yoon's model with a metallicity independent channel through which GRBs can form, with a parameter $p$ controlling the strength of this second channel. It was found that, to best fit the data, a considerable fraction ($10-25\%$) of GRBs should originate from this metallicity independent channel.

Finally, when we measure the metallicity from a distant galaxy, the quantity measured is necessarily a weighted average.\footnote{Different observational techniques will lead to different ways this average is weighted - see Section \ref{subsec:metallicities} for more details.} For this reason, a galaxy with a high measured metallicity may still have some low-metallicity, high SFR regions that could be active sites for GRB formation \citep{Abs-vs-Emiss-Review, Niino+17}. Similarly, a bright, high-metallicity galaxy in close proximity to a neighbouring low-metallicity satellite galaxy could be misidentified as a GRB host.

Studies of spatially resolved star forming regions are possible in the local Universe, and some results have been obtained for GRB hosts at low redshift. Interestingly, for local GRB host galaxies with resolved internal metallicity distributions, only small variances in metallicity are observed. \citet{Kruhler+17} found that GRB980425 at $z=0.0087$ originated from a region with metallicity $\sim 0.1$ dex lower than the metallicity measured from a galaxy-averaged integrated light spectrum.  Similarly, \citet{Izzo+17} found that the site of GRB100316D at $z=0.0591$ had the lowest metallicity, the highest star formation rate, and the youngest stellar population observed in the whole host galaxy. In both of these studies, the host was observed to have a relatively uniform metallicity to $\lesssim 0.3$ dex. These two hosts are low-mass, low-metallicity galaxies in the local Universe, so open questions remain as to (1) whether the very small sample of GRB hosts for which the internal metallicity can be resolved is representative of the general GRB host population, and (2) whether typical GRB host galaxies at higher redshift have the same level of chemical inhomogeneity as their local Universe counterparts. In particular, the latter question might have a negative answer because the galaxy merger rate is approximately constant per unit redshift \citep{fakhouri2010}, hence high-redshift galaxies are more likely to experience infall of chemically pristine clumps of gas. 


At higher redshifts, detailed metallicity maps of GRB host galaxies cannot be resolved. However, the impact of these details can still be quantified using simulations. Combining cosmological hydrodynamical simulations of galaxy formation with a Monte Carlo code to generate GRB events, \citet{Nuza+07} highlighted that GRB host galaxies could still have high metallicities on average due to the complex structure of the interstellar medium, even when GRB progenitors are restricted to be low-metallicity massive stars. More recently, \citet{Bignone+17} used the Illustris simulation to explore the effect of the intrinsic metallicity distribution of galaxies on the stellar mass and metallicity distributions of GRB hosts, using a step-wise metallicity cutoff function, with thresholds from $0.1Z_\odot$ to $ 1.0Z_\odot$, finding that observations of GRBs and their hosts were most consistent with a metallicity cutoff between $0.3Z_\odot$ and $0.6Z_\odot$.

In this paper, we extend on the \citet{Bignone+17} analysis, improving it on several aspects. First, we test a wider range of idealized and realistic metallicity bias functions, including the collapsar model \citep{Yoon+06}, the binary synthesis models of \citet{Chrimes+2020, Eldridge+19}, and the dual-channel model of \citet{Trenti+15}. Second, we take advantage of the availability of simulation data from IllustrisTNG \citep{TNG1}, the state-of-the-art sequel to Illustris. Third, we extend the data-model comparison to include absorption metallicity measurements from the GRB afterglows from \citet{Cucchiara+15} in addition to stellar masses and GRB rate across redshift, carefully undertaking a detailed and accurate data-model comparison.

This paper is organised as follows. Section \ref{sec:methods} gives a brief overview of the IllustrisTNG simulation and describes the process used to determine the likelihood of each galaxy hosting a GRB. In Section \ref{sec:analysis}, we compare the simulated distributions of GRB host masses and metallicities to observational data, and compute the relative likelihoods of the models tested. The mass, metallicity, and redshift distributions of the simulated GRB host population for our most likely model are presented in Section \ref{sec:results}. Discussion and conclusions are presented in Sections \ref{sec:discussion} and \ref{sec:conclusions}.

\section{Methods} \label{sec:methods}

\subsection{The IllustrisTNG simulation} \label{subsec:sim}


The IllustrisTNG simulation is a large volume, cosmological, gravo-magnetohydrodynamical simulation run using the moving-mesh code AREPO \citep{TNG1, TNG2, TNG3, TNG4, TNG5}. It is the follow-up of the Illustris simulation \citep{Illustris1, Illustris2}, containing updated physical descriptions of galactic winds, AGN feedback, chemical enrichment from stars, metal advection, and black-hole accretion, as well as using improved numerical methods. Unlike Illustris, IllustrisTNG also includes magnetism, overall providing improved descriptions of a large set of observations across redshift and galaxy types. 

In this paper, we use the TNG100-1 simulation, in which a $106.5$ cMpc$^3$ box is evolved from a redshift of $z=127$ down to $z=0$. This simulation contains $1.2\times10^{10}$ baryon and dark matter particles, allowing galaxies to be resolved down to length scales of $100$ physical pc within their star-forming regions. The large volume of this simulation ensures that a statistically significant sample of galaxies of different sizes is captured with minimal cosmic variance \citep{trenti08}, while its high resolution allows us to consider the internal metallicity distribution of these galaxies.

\begin{figure*}
\centering

\includegraphics[width=\textwidth]{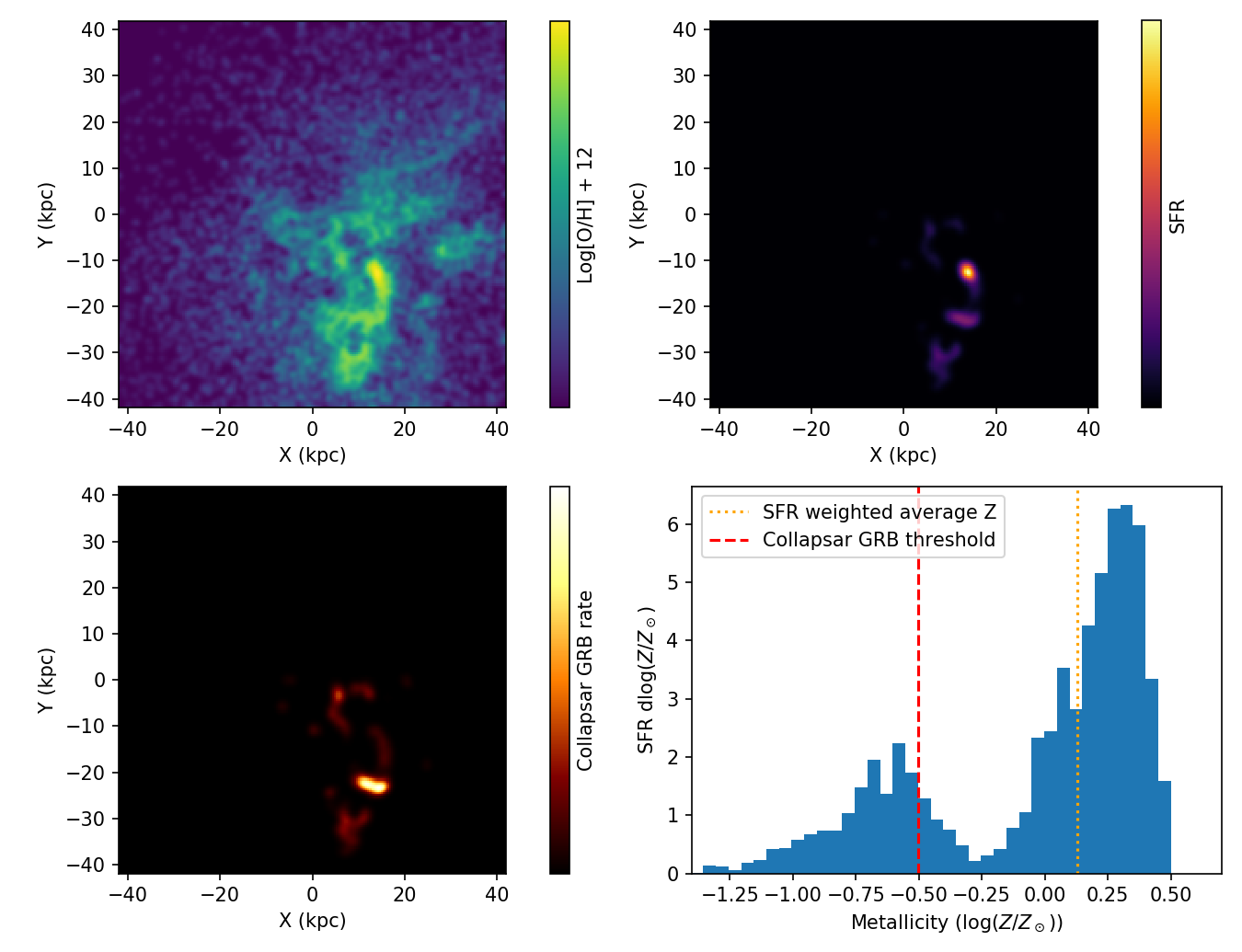}

\caption{The metallicity (top left), SFR (top right), and GRB activity (bottom left) profile for Subhalo \#464361  of the TNG100-1 simulation, taken from Snapshot \#57 at $z=0.76$. This subhalo has a stellar mass of $2.6 \times 10^{9}M_{\odot}$, a star formation rate of $1.93 \, M_\odot/yr$, and a SFR weighted average metallicity of $1.3Z_\odot$. On the bottom right we show a histogram comparing the metallicity of all cells in this galaxy, weighted by their SFR, to the galaxy average (dotted orange vertical line). We also plot the threshold metallicity for GRB production through the collapsar channel, at $Z=0.3Z_\odot$ (dashed red vertical line).
While the SFR-weighted average metallicity of this galaxy is supersolar, it still contains enough star formation activity in low metallicity regions that a collapsar GRB could be hosted with non-zero probability.}
\label{fig:internal_Z};
\end{figure*}

As an illustration, Figure \ref{fig:internal_Z} shows the internal metallicity structure for a small starburst galaxy (${M_* = 2.6\times 10^{9} M_\odot}$) at $z=0.76$. The overall SFR for this galaxy is $1.93\, M_\odot$ yr$^{-1}$, and its average metallicity weighted by SFR is $1.3 Z_\odot$. However, a significant fraction of star formation is occurring in gas cells outside the high metallicity galactic centre, where metallicities are as low as $\approx 0.3 Z_\odot$. Since current gas-phase emission-line spectroscopy techniques can generally only determine a global metallicity measurement for such high-$z$ galaxies, these low-metallicity regions would be missed observationally.
Although galaxies with high global metallicity are not typical GRB hosts, this example shows that
it is possible for galaxies with supersolar metallicities to contain a population of young low-metallicity stars, which would be potential GRB progenitors. Here, we setup a modeling framework to investigate this possibility quantitatively. 


\subsection{Identification of galaxies in the IllustrisTNG simulation} \label{subsec:connect_subhalos}
For each snapshot of the IllustrisTNG simulation, we resort to the publicly available dark matter halo catalogs (inclusive of subhalos), generated using a friends-of-friends (FOF) algorithm \citep{FOF}. To ensure that our data-model comparison is robust, we define a \emph{galaxy} in the IllustrisTNG simulation to be the set of baryonic matter particles (gas+stars) inside subhalos that would not be individually resolved by typical observations. This definition allows us to account and correct for the fact that some subhalos in the public catalogs may represent high-density disk instabilities, rather than satellite galaxies of cosmological origin \citep{TNG2}. 

Four different criteria for the angular resolution of distant galaxies were trialled to determine which subhalos are observationally distinguishable. The simplest is that the half star radii of subhalos in $3D$ space do not overlap; any pair of subhalos with overlapping stellar envelopes may be considered the same galaxy. A more observation-motivated criterion is that the half star radii of subhalos in $2D$ space projected along the line-of-sight must not overlap in order for them to be distinguishable. For ease of computation, the line of sight to any galaxy in this simulation is considered to be along the $z-$axis. This is the most conservative criterion, and will lead to the highest number of connected subhalos. Of course, some subhalos whose stellar envelopes overlap over a line of sight may still be separated based on their relative redshifts or colors. This effect is captured in the third criterion, which states that two subhalos with overlapping half-star radii in $2D$ space must also have a redshift separation of $\Delta z/z <0.002$. Finally, we compare these criteria with the assumption that all subhalos are distinguishable galaxies.

We find that even with our most conservative criteria (Criterion 2), only half a percent of subhalos were found to be connected (0.005) at a redshift $z=2.58$ (Snapshot \#28). The statistical distributions of GRB host properties, such as metallicity and stellar mass, are therefore not affected significantly by the choice of connection criteria. For this reason, we use the simplest criteria for the rest of this analysis - that two subhalos belong to different galaxies if and only if their half star radii do not overlap in 3D space.



\subsection{Metallicity: GRBs vs host galaxies}
\label{subsec:metallicities}

Each galaxy in the simulation contains between $10^2$ and $10^6$ Voronoi-cell gas particles, with cell sizes in the range $10^{-2}-10^{2}$ kpc, depending on the density of gas. The metallicity of each gas particle is defined to be the ratio:
\begin{equation}\label{eq:zcell}
Z_{cell} := M_{Z}/M_{cell},
\end{equation}
where $M_{Z}$ is the mass of all elements heavier than Helium within the Voronoi cell, and $M_{cell}$ is the total gas cell mass. Following the IllustrisTNG simulation public catalog website, we assume the value of Solar metallicity to be $Z_\odot = 0.0127$ (see \url{http://www.tng-project.org/data/docs/specifications/}).

The total metallicity of a galaxy can be defined in several ways. For our purposes, we use definitions aimed at matching the different observational metallicity measurements used for GRB hosts. 
As a proxy for gas-phase emission-line spectroscopy of the UV/optical lines employed at high redshift we weight the cell metallicity by the SFR as proxy for the cell light:  

\begin{equation} \label{eq:z_emiss}
    Z_{emiss} := \frac{\sum_{\text{all gas cells}} Z_{cell} \times SFR_{cell}}{\sum_{\text{all gas cells}}SFR_{cell}}
\end{equation}

Emission-line metallicity measurements for GRB hosts are challenging at high redshift, and available only for a limited number of objects \citep{Kruhler+15,Abs-vs-Emiss-Review}. Thus, we also consider the  sample of $N=55$ absorption-line metallicity determinations from the GRB afterglow presented in \citet{Cucchiara+15}. This measurement is not weighted by the stellar light, but rather by the gas column density along a line-of-sight originating at the GRB's location.

Using the detection of absorption-line variability, \citet{Vreeswijk+2012} estimate that GRBs will ionise the interstellar medium of galaxies starting from approximately 100pc from the site of the burst. To model the metallicity of a GRB host galaxy as observed using the absorption method in the IllustrisTNG simulation, we first randomly select a site from which the GRB originates (see Section \ref{subsec:GRBpos}) and a direction for the line-of-sight (LOS) along which the burst travels. We then identify all gas cells in the galaxy that intersect with this line-of-sight, and compute the length of the section of the LOS that passes through each gas cell. 
Assuming that the LOS has a uniform cross-section throughout the galaxy, we may then determine the fraction of metals along this LOS using the following equation:
\begin{equation}
    \label{eq:Z_DLA}
     Z_{abs} := \frac{\sum_{\text{gas cells along LOS}} Z_{cell} \times \rho_{cell} \times l_{cell}}{\sum_{\text{gas cells along LOS}} \rho_{cell} \times l_{cell}},
\end{equation}
where $l_{cell}$ is the length of the line-of-sight passing through that gas cell.


\subsection{GRB formation models} \label{subsec:models}

The rate of GRBs is proportional to the rate of supernovae, which is proportional to the rate of star formation \mbox{\citep{Fruchter+06}}. The fraction of supernovae in a galaxy that are GRBs depends in some way on the metallicity of the galaxy. For this reason, at least at low redshift, GRBs act as biased tracers of star formation \citep{Hunt+14}.
Combining these facts, we get the following equation for the GRB rate in a star-forming region:
\begin{equation} \label{eq:rate}
    \rho^{(GRB)} = \kappa(Z)\times SFR,
\end{equation}
where $\kappa(Z)$ is the metallicity bias function of GRBs, and $Z$ is the metallicity of the region out of which the GRB forms. 
A range of different models for $\kappa(Z)$ have been proposed (see Subsections \ref{subsubsec:cutoff}-\ref{subsubsec:BPASS}).

We apply Equation~\ref{eq:rate} to compute the rate of GRBs in a galaxy for each metallicity bias model in two different ways. Firstly, we account for the full metallicity distribution of gas cells from the IllustrisTNG simulation by calculating the GRB rate for each cell, and then summing those rates to determine the total GRB rate for that galaxy:
\begin{equation} \label{eq:cell_wise_rate}
    \rho^{(GRB)}_{\text{cell-based}} = \sum_{\text{all gas cells}} \kappa(Z_{cell})\times SFR_{cell}.
\end{equation}

For comparison to previous work, we also consider metallicity as a global galaxy property, and thus define:
\begin{equation} \label{eq:galaxy_wise_rate}
    \rho^{(GRB)}_{\text{unif}} = \kappa(Z_{emiss})\times \sum_{\text{all gas cells}}  SFR_{cell},
\end{equation}
where $Z_{emiss}$ is defined in Equation \ref{eq:z_emiss}. This gives the rate of GRBs for a galaxy, under the assumption that it has a uniform metallicity throughout. 

\subsubsection{Cutoff bias function} \label{subsubsec:cutoff}

The simplest metallicity bias function we test is a stepwise cutoff function. This model contains two free parameters - $\kappa_0$, which describes the total frequency at which GRBs are observed in the universe, and $Z_{max}$, the maximum metallicity that a star may have in order to form a GRB:
\begin{equation} \label{eq:cutoff_bias}
    \kappa_{\text{cutoff}}(Z)=\begin{cases}
      \kappa_{0}, & \text{if}\ Z < Z_{max} \\
      0, & \text{otherwise}
    \end{cases}
\end{equation}

\subsubsection{Two-channel bias function} \label{subsubsec:trenti}

Our second metallicity bias function comes from \citet{trenti13,Trenti+15}. This model is based on the model developed by \citet{Yoon+06}, in which GRBs form from a single rapidly rotating collapsar. \citet{trenti13} proposed that, while some GRBs may originate from this channel, a considerable fraction of them may originate from other channels with no metallicity dependency. 
To codify this theory, this model contains two free parameters: $\kappa_0$, which again describes the total frequency at which GRBs are observed in the universe, and $p$, which controls the relative frequency of GRB production through the metallicity free channel, as compared to the metallicity dependant channel:
\begin{equation} \label{eq:trenti_bias}
    \kappa_{\text{2channel}}(Z)=\kappa_{0} \times \frac{a \log _{10} Z / Z_{\odot}+b+p}{1+p}
\end{equation}
Here, \(a,\) and \(b\) take the same values as chosen in \citet{trenti13}: 
for \(Z / Z_{\odot} \leq 10^{-3}, a=0, b=1 ;\) for \(10^{-3} \leq Z / Z_{\odot} \leq\)
\(10^{-1}, a=-3 / 8, b=-1 / 8 ;\) for \(10^{-1} \leq Z / Z_{\odot} \leq 1, a=-1 / 4\)
\(b=0 ;\) and for \(Z / Z_{\odot}>1, a=0, b=0 .\) These are not free parameters, but rather values representing an interpretation of the \citet{Yoon+06} discrete sampling of metallicity that allows GRBs to form from progenitors with metallicities up to $1.0 Z_\odot$ with a vanishingly small probability. 

\subsubsection{BPASS bias functions} \label{subsubsec:BPASS}

Most massive stars exist in binaries \citep{mostly-binaries}. For this reason, it is likely that most GRB progenitors forming via the single star channel described by \citet{Yoon+06} will still have a companion. Using the binary population and spectral synthesis code BPASS (see \citealt{Eldridge+17} for an overview), \citet{Eldridge+19} computed the rates of GRBs originating from the collapsar channel as a function of redshift for binary stars. They found good agreement between the predicted evolution of the rate of GRBs with redshift and the observations of \citet{SHOALS1}. 

Extending on this analysis, \citet{Chrimes+2020} predicted the metallicity dependence of this formation pathway, and studied how this changes when tidal interactions were included, allowing the high rotational velocities of the primary star necessary for the formation of a GRB to be maintained even in higher metallicity systems (up to $Z \gtrsim Z_\odot$). They found that this model could reproduce the metallicity distribution of GRB host galaxies as observed by \citet{Graham+19, Palmerio+2019, Modjaz+19} and \citet{Japelj+18}; however, in their analysis, the internal metallicity distribution of the host galaxies was ignored. 

To see how their results change when the inhomogeneity of the metal distribution of galaxies is accounted for, we include both of these models in our analysis. The metallicity dependence for both the collapsar model and the binary model powered by tidal interactions is given in Figure 15 of \citet{Chrimes+2020}. Using a plot digitiser, the metallicity bias functions for each of these models were determined. These two metallicity bias functions were also tested in this study.

\subsection{GRB host and position sampling}\label{subsec:GRBpos}






In order to determine $Z_{abs}$ for GRB host galaxies (Equation \ref{eq:Z_DLA}), the precise location of the GRB in the host galaxy must be determined. For this, we resort to Monte Carlo sampling of the probability of forming a GRB in individual computational cells. The probability of a GRB occurring in any gas cell of a given host galaxy is given by:

\begin{equation}
    \label{eq:pr_host_cell}
    Pr^{(GRB)} = \frac{\rho^{\text(GRB)}_\text{cell}}{ \sum_{\text{all gas cells}}\rho^{\text(GRB)}_\text{cell}}.
\end{equation}

This equation follows from Equation \ref{eq:rate} together with Bayes' theorem, and is true regardless of the underlying metallicity bias function. When the metallicity of the galaxy is assumed to be uniform throughout, $Pr^{(GRB)}$ becomes proportional to the SFR of each gas cell, with no dependence on that cell's metallicity.

\section{Results} \label{sec:analysis}

Ten different snapshots from the IllustrisTNG simulation were downloaded, with redshifts ranging from $0.3$ to $9$.\footnote{Snapshots $6, 11, 16, 21, 25, 30, 36, 44, 57$ and $78$ - all available for download at \url{http://www.tng-project.org/data/}. } 
For each snapshot, all galaxies were identified as described in Section \ref{subsec:connect_subhalos}. For each of these galaxies, their GRB rate was calculated for the cell-based and uniform-based approach through Equations \ref{eq:cell_wise_rate}-\ref{eq:galaxy_wise_rate}. A wide selection of metallicity bias functions $\kappa(Z)$ were used, to explore a large area of parameter space for each model. Twenty different cutoff bias functions (Equation \ref{eq:cutoff_bias}) were trialled with cutoffs ranging from $0.05 Z_\odot$ to $1.0 Z_\odot$, increasing in increments of $0.05 Z_\odot$. A large selection of two-channel metallicity bias functions (Equation \ref{eq:trenti_bias}) were also trialled, with $p$ ranging from $0$ to $0.2$ in increments of $0.01$. The binary progenitor models from \citet{Chrimes+2020} and \citet{Eldridge+19} were also included. 

For each GRB-host galaxy, $Z_{abs}$ was computed using each metallicity bias function (see Section \ref{subsec:metallicities}). Using our simulated GRB host population, we then computed distributions of the stellar mass $M_*$ and metallicity $Z_{abs}$, and investigated how these distributions changed with $z$.

In this section, we compare our different metallicity bias models, taking into consideration three key quantities: (1) GRB rate as a function of redshift; (2) host metallicity; and (3) host stellar mass. 

The likelihood of the GRB rate through cosmic time for each model is assessed against the rate derived by the SHOALS survey by computing the value of $\chi^2$ for each model, as described in Section \ref{subsec:GRB_rate}. The likelihood of stellar masses and metallicities for GRB hosts for our models is evaluated against the observations by taking into account both model probability distributions and observational uncertainties by using the law of total probability.

Let $x$ be any observable quantity. Let $p_{sim}(x)$ be the probability density function of $x$ determined from a given theory/numerical model. Then, the probability of measuring $x$ assuming that the model is correct is given by:
\begin{equation}
    \label{eq:general_likelihood}
    Pr(\text{measure } x\, | \text{ sim}) = \newline \int_{-\infty}^\infty Pr(\text{measure } x\,|\,y) p_{sim}(y) dy.
\end{equation}
Here, $Pr(\text{measure } x\,|\,y)$ represents the probability of measuring a value of $x$ for our observable given that the true value is $y$. For example, if we measure $x$ to be greater than a $2\sigma$ lower limit of $x_{min}$, we have that:
\begin{equation}
    Pr(\text{measure } x\,|\,y) = \begin{cases}
    0.95 & \text{if } y \geq x_{min}\\
    0.05 & \text{if } y < x_{min}
    \end{cases}
\end{equation}
If instead we measure a value of $x$ with an uncertainty of $\sigma$, then we model $Pr(\text{measure } x\,|\,y)$ as a standard normal distribution, with mean $x$ and standard deviation of $\sigma$. Thus, Equation \ref{eq:general_likelihood} gives a powerful and versatile way to compare simulations to observations, taking into account the variation in the masses and metallicities of our simulated population of GRBs, and also the uncertainty in our observations. Note that in the special case that the measured value of $x$ has no uncertainty, $Pr(\text{measure } x\,|\,y)$ is a Dirac-$\delta$ distribution, and the likelihood $Pr(\text{measure } x\,| \,\text{sim})$ reduces to $p_{sim}(x)$.

For these calculations of model likelihood, each GRB host observation was matched to the TNG snapshot with the closest redshift. The GRB-DLA metallicity measurements from \citet{Cucchiara+15} are given as either $3\sigma$ lower limits, or observations with $1\sigma$ errorbars. 

For the stellar mass distribution of GRB host galaxies, the BAT6 complete sample of GRB host galaxies was used \citep{Vergani+15, Palmerio+2019}. The stellar masses of 24 galaxies in this sample up to a redshift of $z=2$ were determined by SED fitting. Data quality/availability for 5 galaxies from this survey prevented the authors to carry out an SED fit, and stellar masses were instead estimated from their NIR magnitudes \citep{SHOALS2}. Because there is a tendency for stellar masses determined in the latter way to be overestimated \citep{Ilbert+10}, we treat these values as upper limits, and do not include them when computing the log likelihoods of the various GRB formation models \citep{Palmerio+2019}.

For each model, the likelihood of making each measurement given the model is correct was computed using Equation \ref{eq:general_likelihood}. Because each measurement is independent of all other measurements, the total log likelihoods for observing the mass and metallicity distributions given each of our models
was then calculated by adding together the individual log likelihoods.

\begin{figure*}
    \centering
    \includegraphics[width=0.47\textwidth]{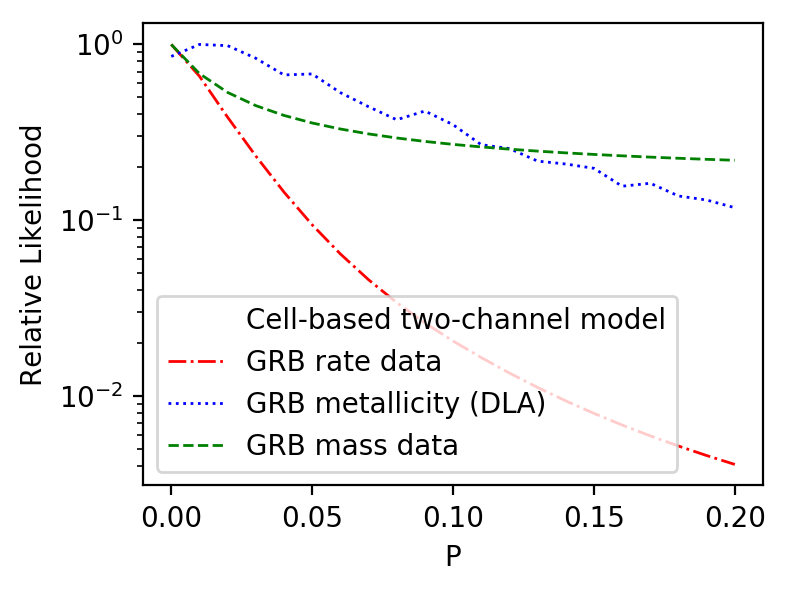}
    \includegraphics[width=0.47\textwidth]{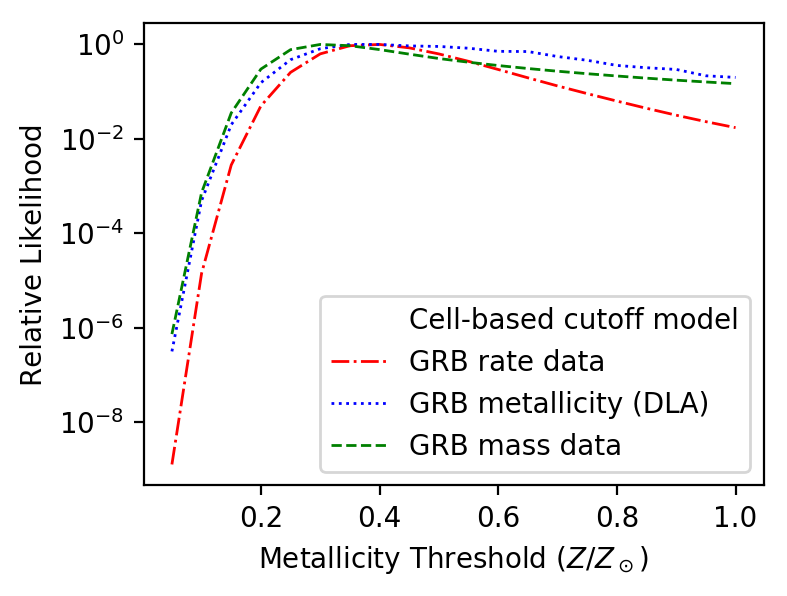}
    \includegraphics[width=0.47\textwidth]{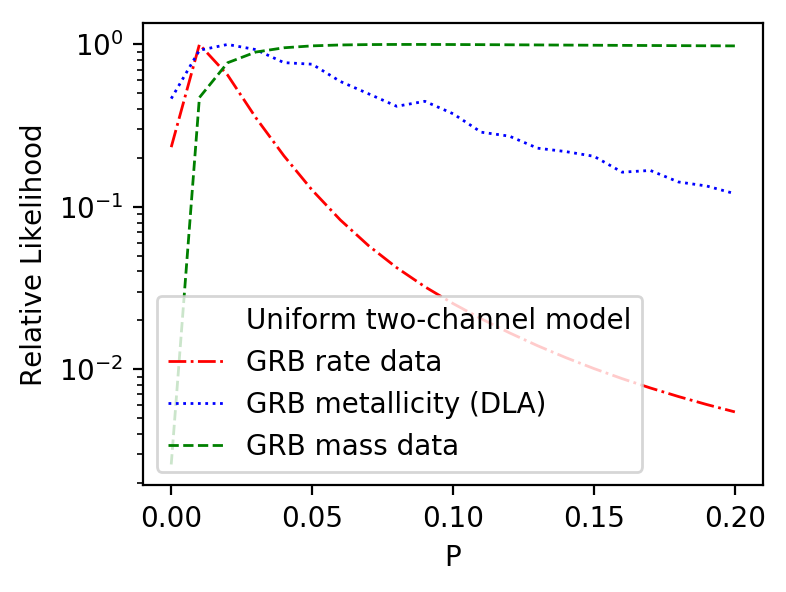}
    \includegraphics[width=0.47\textwidth]{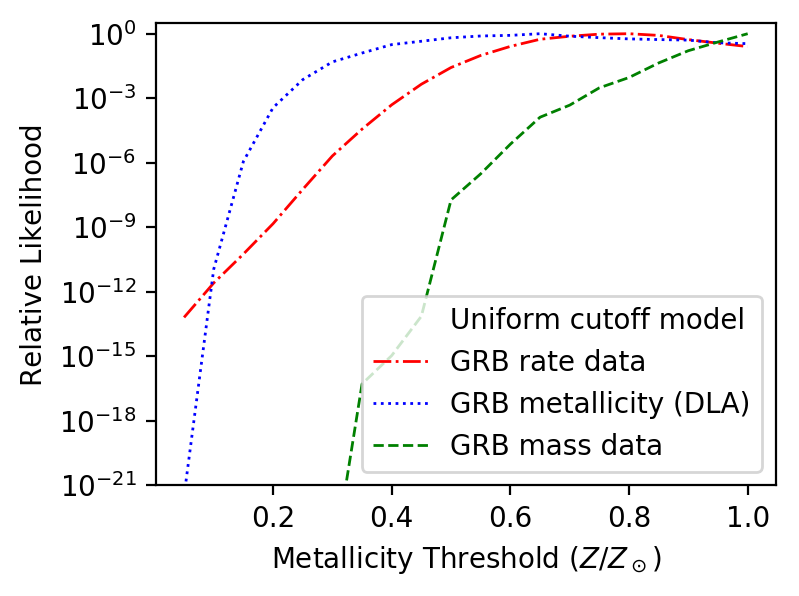}
    \caption{Relative likelihood plots for different models considered in this paper. Dash-dotted red lines are derived from fitting the rate of GRBs through cosmic time to data from \citet{SHOALS1}. Dotted blue lines indicate relative likelihoods derived using metallicity data from \citet{Cucchiara+15}. Dashed green lines indicate relative likelihoods derived using data on the stellar mass of GRB hosts \citep{Vergani+15, Palmerio+2019}. \emph{Top left:} Likelihoods for the two-channel metallicity bias function, accounting for the differing metallicity of each gas cell. \emph{Top right:} Likelihoods for a basic cutoff model, again with cell-based metallicity. \emph{Bottom left:} As in the top left panel, but here the two-channel metallicity bias function is implemented assuming a constant metallicity for each galaxy. \emph{Bottom right:} As in the top right panel, but with a constant metallicity for each galaxy.}
    \label{fig:likelihoods}
\end{figure*}

Figure \ref{fig:likelihoods} shows the relative likelihoods of each of the parameter-dependant $Z$-bias functions for each of our data sets. The top panels show results obtained from models that take into account the metallicity and star formation rates of individual gas cells for each galaxy, whereas the bottom panels show the relative likelihoods computed when the GRB formation models were calculated using only the galaxy average properties. Likelihood values for the best fitting models as well as the two BPASS models are reported in Table~\ref{tab:likelihoods}, using both the inhomogeneous  metallicity distributions of the IllustrisTNG galaxies, and the assumption that these galaxies have uniform metallicity throughout.

\begin{table*}
    \centering
    \begin{tabular}{|c|c|c|}
         \hline
          \textbf{Model} & \textbf{Parameter of best fit} & \textbf{Relative likelihood}  \\
         \hline
         Cell-based cutoff bias & $Z_{max} = 0.35 Z_\odot$ & $1.0$ \\
         Cell-based tidal binary & N/A & $0.44$ \\
         Cell-based two-channel bias & $p=0$ & $0.24$ \\
         Cell-based binary collapsar & N/A & $0.23$ \\
         Uniform two-channel bias & $p=0.02$ & $8.6\times10^{-3}$ \\
         Uniform cutoff bias & $Z_{max} = 1.0 Z_\odot$ & $4.4\times10^{-3}$ \\
         Uniform tidal binary  & N/A & $3.7\times10^{-6}$ \\
         Uniform binary collapsar & N/A & $1.1\times10^{-11}$ \\
         \hline
    \end{tabular}
    \caption{Parameters of best fit for each model, and the relative likelihoods of each model. The metallicity bias function that best matches the data is a cell-based cutoff function with $Z_{max} = 0.35Z_\odot$.}
    \label{tab:likelihoods}
\end{table*}

First, the data model comparison shows that a simple cutoff function with $Z_{max}=0.35Z_\odot$ is overall providing the best match to the current data, although the statistical significance of the result is still limited (i.e. other models cannot be ruled out to high confidence given the likelihood ratios that are reported in Table~\ref{tab:likelihoods}). When a simple cutoff function for the GRB efficiency is used, we find that when the GRB rate of the simulated galaxies is calculated using \mbox{Equation \ref{eq:galaxy_wise_rate}} (uniform metallicity across the whole host), then the best match between the simulation and the GRB rate data occurs at a metallicity threshold $Z_{max} = 0.8Z_{\odot}$, the best match with the metallicity data occurs when $Z_{max} = 0.65Z_{\odot}$, and the best match with the stellar mass data occurs when $Z_{max} > Z_{\odot}$.
However, when the internal metallicity distribution of these simulated galaxies is accounted for, we find that the most likely cutoff from the GRB rate and metallicity data is $Z_{max} = 0.4Z_{\odot}$, and from the mass data is $Z_{max} = 0.3Z_{\odot}$ - that is, all three independent observations have their peak likelihoods at  $\sim0.35Z_\odot$. Furthermore, at every value of $Z_{max}$ tested, we find that the cutoff model is significantly more likely when Equation \ref{eq:cell_wise_rate} is used to determine the rate of GRBs for each galaxy - that is, when the internal metallicity distribution of GRB host galaxies is accounted for.

When the IllustrisTNG galaxies are treated as though they have uniform metallicities throughout, our analysis strongly disfavours both of the BPASS models. This is mainly due to a tension between the predicted and observed mass distributions of GRB hosts for these two models - both of these models fail to explain the population of higher-mass GRB hosts at intermediate redshifts ($1.5\lesssim z \lesssim 2$).
However, when the internal metallicity distribution of the host galaxies is considered, these two models become very competitive. When tidal interactions are ignored, the binary collapsar model is equally likely as the single star collapsar model (the two-channel bias function with a value of $p=0$). This makes sense, since both models are based on the work of \citet{Yoon+06}. We find that the model that includes tidal interactions is twice as likely as these two models, indicating that tidal interactions in close binaries may be a factor in the production of at least some GRBs.

Turning to the two-channel model, when a uniform metallicity for a host galaxy is used, we find that the most likely model from the cosmic GRB rate data and from the metallicity observations is $p=0.02$ (although the mass data shows a preference for models with $p>0.05$), whereas when we consider individual metallicities and star formation rates of gas cells, then both these datasets support the model where $p=0$, i.e. the canonical collapsar model. 

For all models considered, those that best fit the GRB cosmic rate \mbox{\citep{SHOALS1}} tend to broadly agree with those fitting optimally the GRB host mass and metallicity data (\citealt{Cucchiara+15, Vergani+15}, \mbox{\citealt{Palmerio+2019}}). This implies that these three independent data products are roughly in concordance, and can  be explained by the same physical process. 
Finally, we note that a further understanding of the GRB host mass distribution at higher redshifts ($z>2$) would be very powerful to discriminate further between models. This task would be within the capabilities of spectroscopic observations of GRB host galaxies with the James Webb Space Telescope. 

\section{GRB host population statistics for most likely model} \label{sec:results}
 
In this section, we present data products for our most likely $Z$-bias function - the simple cutoff model with a threshold of $Z_{max}=0.35Z_\odot$, and compare how the statistics generated using this bias function compare to observational data.

\subsection{The cosmic GRB rate} \label{subsec:GRB_rate}
\begin{figure}
\centering
\includegraphics[width=0.47\textwidth]{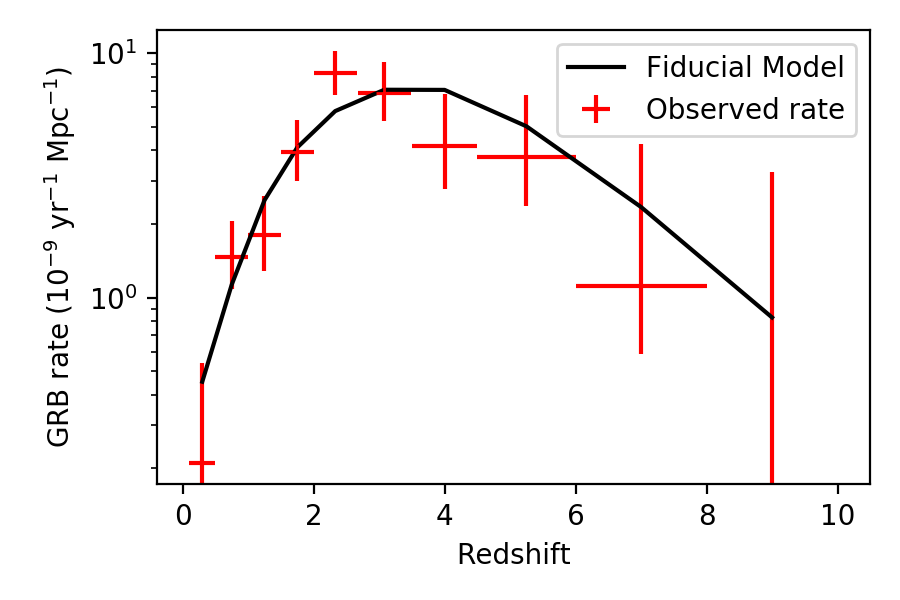}
\caption{Solid line: rate of GRBs as a function of redshift, assuming our fiducial metallicity bias function (cutoff at $Z_{max} = 0.35Z_\odot$) and cell-based metallicity definition (Eq.~\ref{eq:zcell}). For comparison data from \citet{SHOALS1} are shown in red.}
\label{fig:rate_vs_z}
\end{figure}

For each snapshot we derive the total GRB rate, starting from the rate for each galaxy. Because the total comoving volume of the IllustrisTNG simulation is constant with redshift, by summing over all galaxies at each redshift we obtain a theoretical prediction for the rate of GRBs per cMpc$^3$ as a function of $z$. We then compare this statistic to the rate of GRBs per cMpc$^3$ as computed by \citet{SHOALS1} using the complete SHOALS sample of GRBs. We note that this recent determination of the GRB rate is consistent with earlier studies (e.g. \citealt{Wanderman&Piran10, Jakobsson+12, Robertson+Ellis12, Salvaterra+12}) within the respective uncertainties of the analyses. The SHOALS cosmic GRB rate is based on the largest unbiased sample of GRB observations to date and has been determined by carefully taking into account factors that affected observations by Swift and ground and space-based follow-ups to measure the redshift of the GRB afterglow and of its host. As such, we consider it as our state-of-the-art reference, but results of this analysis would not be significantly altered if we were to utilize instead the GRB rate published in earlier studies. We normalise the rate of simulated GRBs by finding the value of $\kappa_0$ that best fits the \citet{SHOALS1} cosmic GRB rate data using $\chi^2$ minimisation. This result is shown in Figure \ref{fig:rate_vs_z}. Our fiducial model fits the data with a $\chi^2$ value of $6.05$, indicating that this cutoff model combined with a cell-based metallicity definition provides a good description of this observable.

\subsection{Absorption metallicity vs emission metallicity for GRB host galaxies} \label{subsec:Z_emiss_vs_Z_DLAs}

\begin{figure}
\centering
\includegraphics[width=0.45\textwidth]{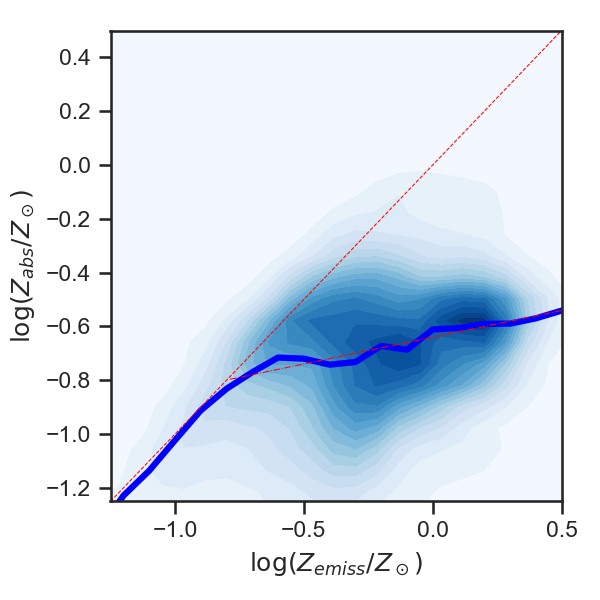}
\caption{Contour plot showing the metallicity as measured by the emission-line method against the metallicity measured by the absorption method for 10,000 GRB hosts at $z=2.33$, assuming our fiducial GRB model (cutoff with $Z_{max} = 0.35Z_\odot$). The bold line plots the median $Z_{abs}$ as a function of $Z_{emiss}$. The red dashed line shows where $Z_{abs}=Z_{emiss}$. At lower metallicities, these two lines are fairly consistent with each other, but for higher metallicity hosts ($Z \gtrsim 0.2Z_\odot$), $Z_{abs}$ becomes significantly lower than $Z_{emiss}$. The red dash-dotted line shows the linear regression model of best fit between $Z_{abs}$ and $Z_{emiss}$ for $\log(Z_{emiss}/Z_\odot) > -0.8$.}
\label{fig:AbsZ_vs_EmissZ}
\end{figure}

To build robust host galaxy predictions, we begin by exploring the difference between the emission and absorption metallicity measurements described in Section \ref{subsec:metallicities}. Using Monte Carlo methods, $10,000$ GRB host galaxies were drawn from our simulation at a redshift of $z=2.33$, the redshift at which GRBs are most frequently observed \citep{SHOALS1}. For each galaxy, both the emission line metallicity and the absorption metallicity were computed. A contour plot comparing the relative distributions of these two metallicities for our fiducial model is shown in Figure \ref{fig:AbsZ_vs_EmissZ}. We also plot the median value of $Z_{abs}$ for each value of $Z_{emiss}$. For lower metallicity galaxies ($Z_{emiss} \lesssim 0.2 Z_\odot$), we find that $Z_{abs}$ and $Z_{emiss}$ are fairly similar, but for higher metallicity galaxies, $Z_{abs}$ is vastly smaller than $Z_{emiss}$.
This general trend is to be expected from our model, because the GRB-DLA method should preferentially probe gas near the low metallicity star forming region where the GRB is created. For a metal-rich host, the line-of-sight of the GRB will seldom pass through the high-metallicity galactic centres.

For galaxies with metallicities greater than $0.2Z_\odot$, we find that the median $\log(Z_{abs} / Z_\odot)$ scales approximately linearly with $\log(Z_{emiss} / Z_\odot)$. Using a simple linear regression model, we find that the line of best fit is $\log(Z_{abs} / Z_\odot) = 0.2\log(Z_{emiss} / Z_\odot)-0.64$, with $r=0.96$. 

This analysis explains the lack of supersolar GRB hosts in the sample observed by \citet{Cucchiara+15} when compared to surveys of GRB host galaxies that use emission-line gas phase metallicity measurements, e.g. \citet{Kruhler+15, Palmerio+2019}. 
Of the 10,000 selected GRB hosts presented in Figure \ref{fig:AbsZ_vs_EmissZ}, 25\% have $Z_{emiss} > Z_\odot$; however, only 0.5\% have $Z_{abs} > Z_\odot$. Because \citet{Cucchiara+15} report precise metallicities for only 16 galaxies, the probability of finding a supersolar host in this sample is extremely low ($0.076$, using a binomial distribution).

We note that for different metallicity bias functions $\kappa(Z)$, we expect the shape of the relationship between $Z_{abs}$ and $Z_{emiss}$ to be different. For a sharper metallicity cutoff, $Z_{abs}$ will flatten out with respect to $Z_{emiss}$ at smaller metallicity values. In the extreme case where GRB formation has no metallicity dependence, we expect $Z_{abs}$ to trace the star formation rate weighted $Z_{emiss}$ at all metallicities. These considerations suggests that if a sufficiently large sample of GRB hosts has both absorption and emission-line metallicity measurements, then their comparison could offer a novel metric for constraining the GRB metallicity bias function. Given the lack of such a sample, this analysis is outside the scope of this work.  

\begin{figure*}
\centering
\includegraphics[width=0.48\textwidth]{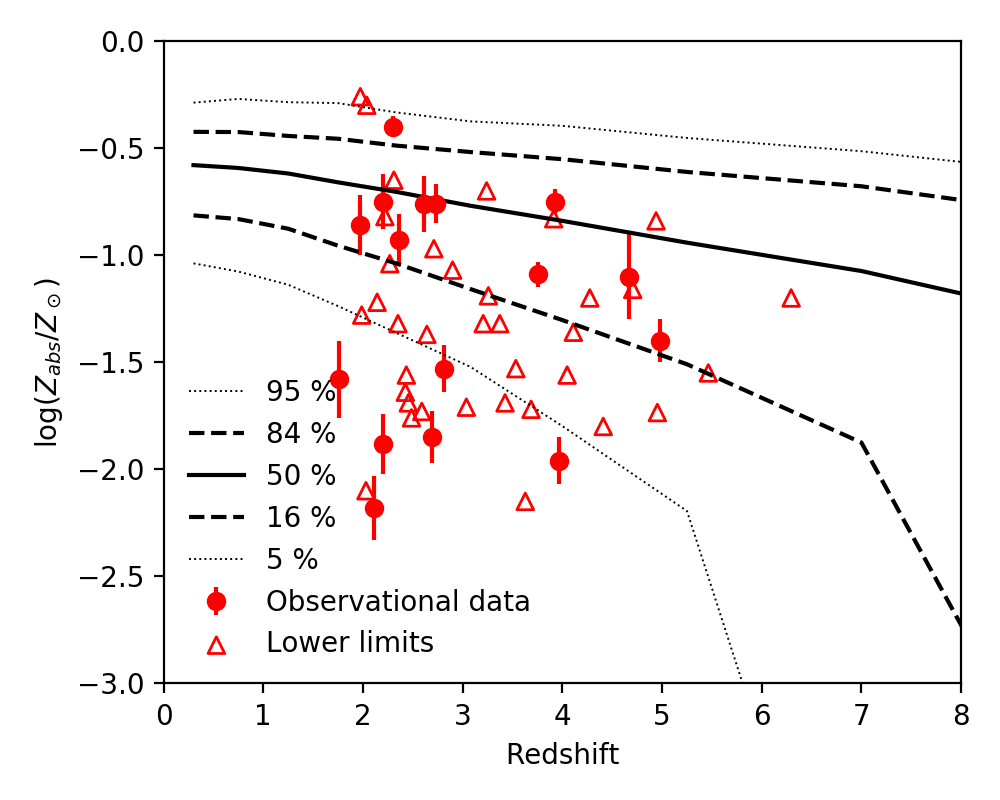}
\includegraphics[width=0.48\textwidth]{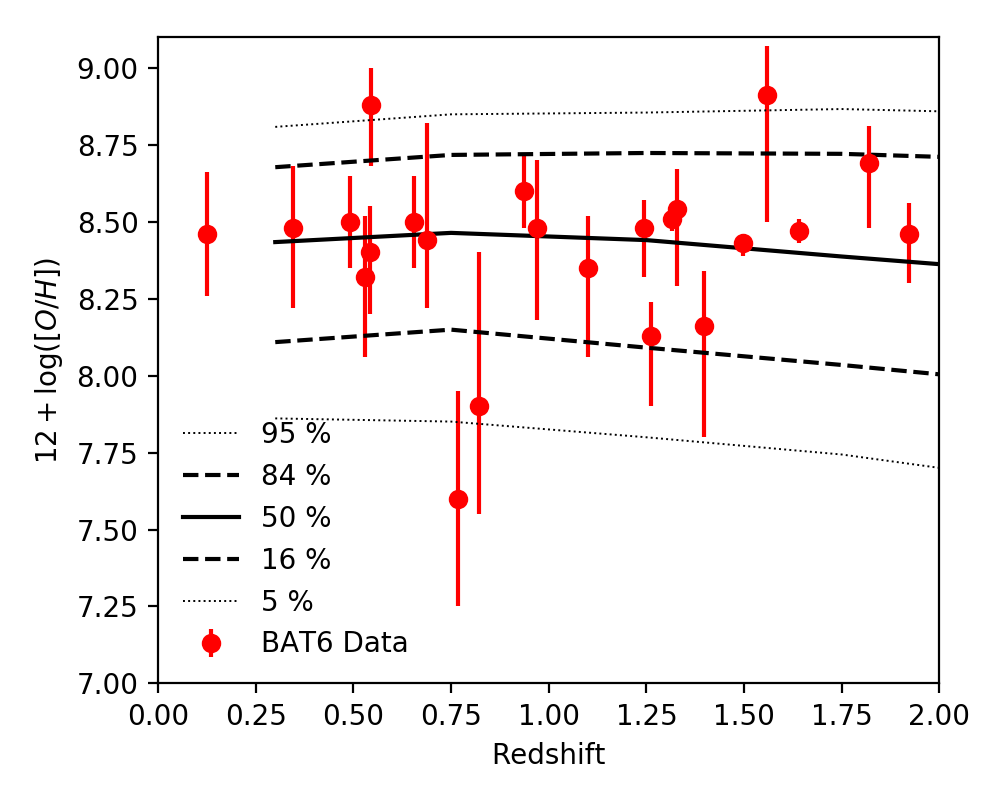}
\caption{\emph{Left:} Black lines show the evolution of the metallicity distribution of GRB host galaxies with redshift, as measured using the absorption method, assuming our fiducial collapsar model (median: solid line; 1 and 2$\sigma$ confidence contours shown as dashed and dotted lines respectively). For comparison we have also plotted metallicity data from \citet{Cucchiara+15}. Red triangles denote $3\sigma$ lower limits. \emph{Right:} As for the left panel but plotting emission-line metallicity of GRB host galaxies (up to $z=2$). Data points from the BAT6 survey \citep{Japelj+16, Palmerio+2019} are overplotted in red to our fiducial model predictions.}
\label{fig:Z_vs_z}
\end{figure*}

\subsection{Metallicity of GRB hosts over cosmic time} \label{subsec:Z_vz_z}

For each snapshot analysed, we construct the absorption and emission metallicity distribution of GRB host galaxies, assuming our fiducial cutoff bias model and using Equation \ref{eq:Z_DLA}. In Figure \ref{fig:Z_vs_z}, we plot the evolution of the absorption metallicity $Z_{abs}$ and emission metallicity $12+\log([O/H])$ of simulated GRB host galaxies against redshift. We also include GRB-DLA metallicity measurement data from \citet{Cucchiara+15}, where triangles are $3\sigma$ lower limits, as well as emission line metallicity measurements of GRB hosts in the BAT6 sample \citep{Japelj+16, Palmerio+2019}. We find that the spread in the metallicity distribution of GRB hosts increases with increasing redshift, and that the median metallicity of GRB hosts decreases slowly with redshift. We find that the predictions of our fiducial model tend to agree very well with the emission line metallicity distrubution of GRB hosts at $z<2$ measured by \citep{Japelj+16, Palmerio+2019}, and are broadly consistent with \mbox{\citet{Cucchiara+15}} data at higher redshifts.

\subsection{Collapsar host galaxies with supersolar metallicity} \label{subsec:supersolar_Z}

At face value, the observation of GRB host galaxies with metallicities greater than $Z_\odot$ cannot be explained by the collapsar model. \citet{Kruhler+15} and \citet{Palmerio+2019} report the fraction of GRB hosts with metallicity greater than solar at $z<1$ to be $\sim 20\%$, decreasing to $\lesssim 10\%$ at $1<z<2$. \citet{Graham+19} also report supersolar hosts at the $\sim 10\%$ level, but do not see evidence for strong redshift evolution out to $z\lesssim 2.5$. In this subsection, we show that a population of supersolar hosts at low redshift can easily be accounted for by our fiducial GRB model, that combines the collapsar metallicity bias with the local cell-based determination of metallicity. 

\begin{figure}
\centering
\includegraphics[width=0.45\textwidth]{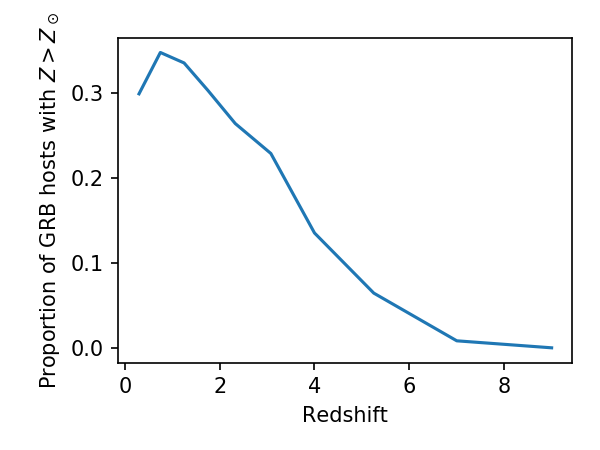}
\caption{Fraction of GRB hosts with $Z_{emiss} > Z_\odot$, according to our fiducial collapsar model. At lower redshifts,  $\sim 30 \%$ of GRB hosts have metallicities above solar, providing a mechanism to solve the tension with observations, without the need to introduce a metallicity-independent GRB formation channel.}
\label{fig:SupersolarHosts}
\end{figure}

While our metallicity bias model prevents the formation of GRBs from stars with supersolar metallicity, it is possible that a supersolar galaxy has a low metallicity region that could contain GRB progenitors (see Figure \ref{fig:internal_Z}). For each snapshot, we compute the proportion of GRB hosts with SFR-weighted average metallicities greater than $Z_\odot$. A plot of the proportion of GRB hosts with metallicities greater than $Z_\odot$ against redshift is shown in Figure \ref{fig:SupersolarHosts}. Here, we see that when the internal metallicity distribution of galaxies is accounted for, the expected fraction of supersolar GRB hosts at $z<1$ is just over 30\%, which is significantly larger than the observational limits determined by \citet{Kruhler+15} and \citet{Palmerio+2019}. This shows that the supersolar GRB hosts observed are not in tension with the collapsar model at all. The fact that our fiducial model predicts even more supersolar hosts than observed could be related to specific details of the chemical enrichment of the IllustrisTNG simulation. In particular, the metallicity of small galaxies in the IllustrisTNG simulation ($\log(M_*/M_\odot) \lesssim 9.5$) is less than the metallicities observed by \citet{MZR_z=0} and \citet{MZR_z=1}; furthermore, the mass-metallicity relation seen in the IllustrisTNG simulation flattens out for galaxies with ($\log(M_*/M_\odot) \lesssim 9.0$) at intermediate redshifts, an effect that has not been observed (see Figure 6 of \citealt{TNG_MZR}).

Figure~\ref{fig:SupersolarHosts} also shows the fraction of supersolar hosts at higher redshift. Overall, we find that 19.4\% of GRBs at $0<z<9$ are hosted by galaxies with supersolar metallicity. This is similar to the fraction of GRB hosts with supersolar metallicity found in \citet{Trenti+15}. Interestingly, their analysis was carried out assuming a single metallicity for GRB host galaxies with the best fit model predicting that 17\% of GRBs should originate through a metallicity independent channel in order to produce the high-metallicity hosts. However, we have shown here that when we account for the internal metallicity distribution of GRB hosts, the observation of supersolar GRB hosts is expected from the collapsar model, removing in principle any need for a metallicity-independent channel (see Section~\ref{sec:analysis} for a quantitative analysis).

\begin{figure*}
\centering
\includegraphics[width=0.8\textwidth]{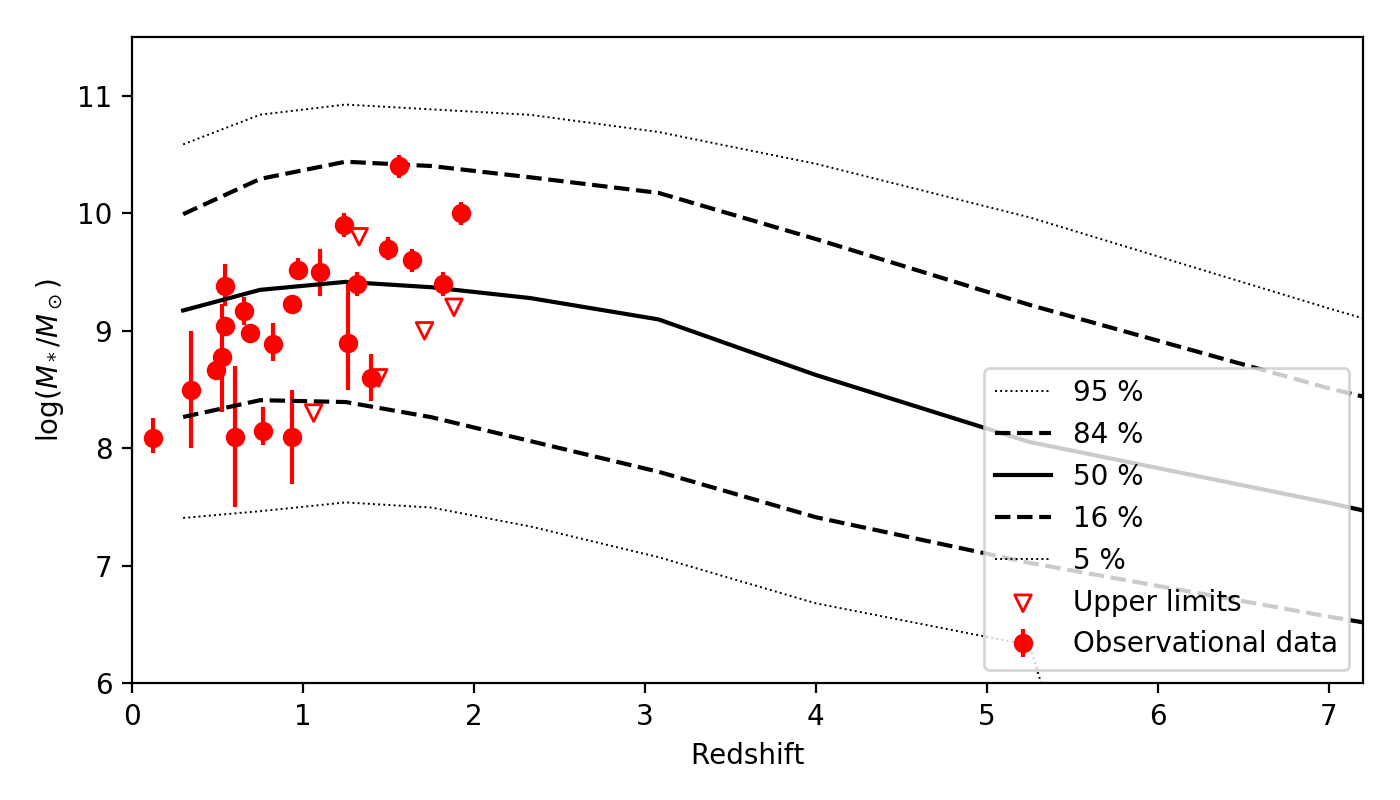}
\caption{The stellar mass distribution of GRB host galaxies with time, using our most likely GRB formation model. For comparison we have also plotted data from \citet{Vergani+15} and \citet{Palmerio+2019}. Red triangles denote upper estimates on GRB host stellar masses.}
\label{fig:M*_vs_z}
\end{figure*}
\subsection{GRB host stellar masses} 

In Figure \ref{fig:M*_vs_z}, we plot the stellar mass distribution of GRB hosts with redshift, according to our best fitting cutoff model. We compare these results with the stellar mass determination for the galaxies in the BAT6 complete sample up to $z=2$. Our model predictions show a slowly increasing median stellar mass out to $z\sim 1.5$ and then a steadily decreasing trend with increasing redshift. This is broadly in agreement with \citet{Jimenez&Piran2013}, who find that most GRB hosts should have stellar masses below $10^{10}M_\odot$; however, it is slightly in tension with the mass estimates of \citet{Vergani+15} and \citet{Palmerio+2019}, which show a steeper increase of median stellar mass with redshift for GRB host galaxies with $0<z<2$. 
Determinations of the stellar masses of GRB host galaxies at higher redshifts will allow us to test our prediction that the masses of GRB host galaxies should decrease with redshift beyond $z \sim 2$.

\section{Discussion} \label{sec:discussion}

From the maximum likelihood analysis presented in \mbox{Section \ref{sec:analysis}}, it appears that when internal metallicity distribution of galaxies is considered, 
a model of GRB formation with a sharp cutoff at $Z_{max} = 0.35Z_\odot$ offers the best match to observations. This threshold is remarkably similar to the metallicity threshold of $0.3Z_\odot$ for the formation of collapsars found by \citet{Yoon+06}.
Interestingly, Table~\ref{tab:likelihoods} shows that when the internal metallicity distribution of GRB host galaxies is ignored, the best fitting model has a small fraction ($p=0.02$) of GRBs forming via a metallicity-independent channel. 
This shows that our study is fully consistent with the previous work of \citet{Trenti+15}, as the difference in outcomes is derived from an improved modeling of host metallicity thanks to the use of the IllustrisTNG snapshots. 

The simulation-based modeling also naturally allows us to consider the scenario in which some GRBs are formed inside a low mass, low metallicity satellite galaxy blended in projection with a larger, brighter, metal-rich galaxy nearby. We find that in IllustrisTNG this scenario, while plausible, is rare enough that it will not significantly affect the statistical distributions of the masses and metallicities of GRB hosts, and so by itself it can not explain the large number of supersolar metallicity GRB host galaxies that have been observed, which originate instead from pockets of low metallicity gas inside  starburst galaxies.

The simple cutoff models shown in Figure~\ref{fig:likelihoods} also allows us to compare our results to previous investigations that have not taken into account the internal metallicity structure of galaxies. Using mass, metallicity, and GRB rate estimates from the SHOALS survey \citep{SHOALS1, SHOALS2}, \citet{Vergani+17} found a best-fitting cutoff of $Z_{\mathrm{max}}=0.73_{-0.07}^{+0.08} Z_{\odot}$. Similarly, using the BAT6 complete sample, \citet{Palmerio+2019}  found that suppression of GRB formation in hosts with metallicities greater than $Z=0.7Z_\odot$ could explain the observational data. When we compare our simulated GRB host population to the rate data of \citet{SHOALS1} and the GRB-DLA metallicity distribution observed by \citet{Cucchiara+15}, we also see that when galaxies are assumed to be of homogeneous metallicity throughout, the most likely explanation is a cutoff function with $Z_{\mathrm{max}}=0.7 Z_\odot$.
However, when the internal metallicity distribution of galaxies is accounted for, the cutoff of best fit drops to $Z_{\mathrm{max}}=0.35Z_\odot$.

In Section \ref{subsec:Z_emiss_vs_Z_DLAs}, it was shown that the observed metallicity of GRB hosts depends on the method used to measure the metallicity. Observations of distant galaxies often rely on emission line spectroscopy techniques, which measure primarily the oxygen content of star forming regions of galaxies. These techniques may lead to overestimations of the iron content of GRB host galaxies \citep{Hashimoto+18}, which is the most important element considered when modelling GRB metallicity bias functions \citep{Yoon+06}.
In this work, we have avoided this issue by comparing the simulation data to observations made using the GRB-DLA method, which probes the content of heavy metals such as Zn, Fe and Si \citep{Cucchiara+15}. 
We found that metallicities measured in this way are significantly lower than those measured using emission-line spectroscopy, especially for galaxies with higher metallicities (see Figure \ref{fig:AbsZ_vs_EmissZ}).

IllustrisTNG has the capability to track the evolution of several chemical species including Fe, Si and O through each gas cell. In future works, this information could be used to determine the rate of GRBs in galaxies based solely on their iron content, and then compare the measured metallicity of these galaxies based on emission line or other methods.

Although our best-fitting metallicity bias function has a metallicity threshold similar to the collapsar model \citep{Yoon+06}, it is not physically motivated. Of the three theory-based models tested (single star collapsar, binary collapsar, binary collapsar with tidal interactions), we find that the most likely model involves the formation of some fraction of GRB progenitors through tidal interactions in binary systems \citep{Chrimes+2020}. In this model, GRB activity peaks at $Z \sim 0.2 Z_\odot$, but there is no cutoff until $Z \sim 1.1 Z_\odot$.
We do not currently have enough data about the population of GRB hosts to constrain the shape of our bias function beyond an estimate of the threshold metallicity for GRB formation.

As remaining point of data-modeling tension, our fiducial collapsar model predicts a relatively large fraction of GRB hosts with supersolar metallicities at low redshifts (see Figure \ref{fig:SupersolarHosts}, Section \ref{subsec:supersolar_Z}). This may suggest an overproduction of metals in simulated low-mass galaxies by the chemical enrichment model of the IllustrisTNG simulation (some tension between the simulation results and the observed mass-metallicity relation at high-redshift is also hinted in the comparison presented in \citealt{TNG_MZR}).

Of course, all our quantitative conclusions depend on the accuracy of the internal metallicity distributions of galaxies in the IllustrisTNG simulation.
Observations of GRB host galaxies in the local universe for which the internal metallicity distribution may be resolved to scales of up to a few 100pc (similar to the scale of star forming gas cells in TNG100-1) show that the site of GRB formation tends to  be $\sim 0.1$ dex lower than the average metallicity of the GRB host \citep{Levesque+11}. Recent studies of the internal metallicity distributions of local GRB host galaxies by \citet{Kruhler+17} and \citet{Izzo+17} both find that GRB occurs in a local environment for which $Z\sim 0.3Z_\odot$, which concurs with our model. 

Observations of the internal metallicity distribution of the host galaxy of the supernova-less long GRB111005A \citep{Tanga+18} found that the burst originated from a site with $Z \approx 0.75 Z_\odot$, with a resolution of $270$ pc. The site of this GRB also showed little ongoing star formation and an old stellar population, limiting the initial mass of the GRB progenitor star(s) to be $<15M_\odot$. This unusual burst may represent the first piece of direct evidence for a metallicity independent GRB formation channel, as predicted by \citet{Trenti+15}. However, due to the edge-on orientation of the galaxy, it is also possible for this burst to originate from a low-metallicity region either in front of or behind the galactic centre. In any case, our model shows that if there is a metallicity-independent GRB formation pathway, then it is unlikely to produce a significant fraction of observed GRBs.

A rigorous comparison of the internal metallicity distributions of galaxies in the IllustrisTNG simulation to those observed has not yet been performed, and is beyond the scope of this work. However, we argue that even if the internal metallicity distributions of these simulated galaxies are not accurate, to assume that galaxies have a uniform metallicity throughout is even less accurate. 
Future studies that model the metallicity bias of GRB progenitors must somehow account for the internal metallicity distributions of galaxies.

\section{Summary and conclusions} \label{sec:conclusions}

The origin of long-duration GRBs and the properties of their host galaxies has been previously studied extensively without reaching consensus on whether these explosions originate from a single progenitor, from binaries, or from a combination of both (see e.g. \citealt{GRB_review} for a review). Progress has been limited by the intrinsically small number of GRBs, by the even smaller sample of hosts whose properties have been characterized, as well as by the use of simplified models of GRB formation that often assume a single metallicity for a galaxy. 

In this work, we investigated the extent to which accounting for the internal metallicity distribution of GRB hosts affects the conclusions drawn from data-model comparison, using the state-of-the-art IllustrisTNG simulation and post-processing its snapshots with several GRB formation models discussed in the literature. 
Our key conclusions are the following: 
\begin{itemize}
    \item Taking into account internal metallicity of a galaxy qualitatively changes the inference, and it is therefore critically important, as recently identified through use of the original Illustris simulation by \citet{Bignone+17}.
    \item Overall, a simple cutoff model with a metallicity of $Z_{max} = 0.35Z_\odot$ is the one with the highest joint likelihood from comparison to the GRB rate versus redshift, GRB host stellar mass distribution, and GRB-DLA metallicity. A metallicity bias of this form would imply that GRBs could be used as (almost) unbiased tracers of the cosmic SFR at high redshift, when the typical metallicities of host galaxies are expected to be below the cutoff threshold. 
    \item However, we cannot rule out models that involve binary interactions. We find that a metallicity bias model that accounts for the formation of GRB progenitors through tidal interactions in binary systems is almost twice as likely as a model that ignores these interactions.
    \item It is important to carry out data-model comparison self-consistently. In particular the absorption and emission line measurements of metallicities are expected to be different for GRB hosts in presence of a metallicity bias. The shape of their relation may hold further diagnostic power to quantify the bias.
    \item Some remaining tension with observational data remain for our implementation of the collapsar model, specifically in the area of predictions for the redshift dependence of the stellar masses of the GRB hosts, as well as for the fraction of high-metallicity host galaxies at low redshift. Reasons for these discrepancies may include an overproduction of metals in low-mass systems in the IllustrisTNG simulation, and possible inaccuracies in the internal metallicity distributions of these simulated galaxies. A full comparison of the metallicity gradients of IllustrisTNG galaxies to those observed, e.g. by \citet{Belfiore+2017}, is necessary to validate our model.
    Also, direct measurements of stellar masses for high redshift GRB hosts, which will become possible with the James Webb Space Telescope (JWST) would be highly beneficial for offering stronger and more robust observational constraints on the origin of long GRBs. 
\end{itemize}

Last, but not least, to unlock the full potential of long-duration GRBs as tools to characterize star formation across time, a larger sample of events, in particular at high-redshift is needed. While recent progress has been limited, the Neil Gehrels Swift Observatory is still operational, and an array of future space telescopes, large and small\footnote{Approved and in-development missions include: JWST, SVOM, THESEUS, the HERMES constellation of CubeSats, and the SkyHopper Space Telescope Cubesat.} will contribute to identifying and characterizing GRBs and their hosts. 

\section*{acknowledgements}
We thank Antonino Cucchiara for useful discussions on GRB-DLA metallicity, and Sean Crosby for his guidance in using the supercomputing cluster SPARTAN. 
We thank the anonymous reviewer for their helpful feedback, which guided the direction of this paper. This research was partially supported by the Australian Research Council Centre of Excellence for All Sky Astrophysics in 3 Dimensions (ASTRO 3D), through project number CE170100013.

\bibliographystyle{mnras}
\bibliography{biblio} 
\label{lastpage}
\end{document}